\documentclass[aps,prl,twocolumn,groupedaddress,showpacs]{revtex4}
\usepackage{graphicx}

\begin{document}

\title{Observation of ``Partial Coherence" in an Aharonov-Bohm
Interferometer with a Quantum Dot}
\author{Hisashi~Aikawa, Kensuke~Kobayashi, Akira~Sano,
Shingo~Katsumoto, and Yasuhiro~Iye}
\affiliation{Institute for Solid State Physics, University of Tokyo, 5-1-5
Kashiwanoha, Chiba 277-8581, Japan}
\date{\today}

\begin{abstract}
We report experiments on the interference through spin states of
electrons in a quantum dot (QD) embedded in an Aharonov-Bohm (AB)
interferometer.  We have picked up a spin-pair state, for which the
environmental conditions are ideally similar.  The AB amplitude is
traced in a range of gate voltage that covers the pair.  The behavior
of the asymmetry in the amplitude around the two Coulomb peaks agrees
with the theoretical prediction that the spin-flip process in a QD is
related to the quantum dephasing of electrons.  These results
constitute evidence of ``partial coherence" due to an entanglement of
spins in the QD and in the interferometer.
\end{abstract}

\maketitle

Mesoscopic systems are excellent test stages of quantum coherence and
decoherence, which has been one of the most significant and
challenging issues both for fundamental physics and for the
realization of quantum devices~\cite{imry}.  In the context of the
``system-plus-environment" model, decoherence of a particular state of
the system occurs through its coupling to infinite degrees of freedom
of the environment~\cite{leggett}.  The environment affects the
interference in two ways: by dissipation of the system's kinetic
energy into the environment and by phase randomization via interaction
with environmental degrees of freedom.

Aharonov-Bohm (AB) interference is a standard mesoscopic tool for
probing the degree of coherence~\cite{AB1963}; most simply, its
amplitude is a good measure of coherence.  In our previous
paper~\cite{Kobayashi2002JPSJ}, we reported that the AB amplitude in a
semiconductor sample is markedly affected by the coupling to the
environment~\cite{Seelig}.  In a hybrid system of an AB ring and a
quantum dot
(QD)~\cite{Yacoby1995PRL,Katsumoto1996JPSJ,Schuster1997Nature,
Buks1998Nature,vdWiel2000Science,Ji,Kobayashi2002PRL,Kobayashi2003PRB},
one would be able to vary the strength of such coupling.  Buks {\it et
al.}~\cite{Buks1998Nature} used a dot in an AB ring as a which-path
detector.  In their setup, the diminishment of the AB amplitude was
not very large, but correlation with shot noise in the current flowing
through a quantum point contact (QPC) placed next to the dot was
detected.  Because it is unlikely that decoherence occurs when the
current meter ``notices" the passage of an electron over the dot from
the shot noise, this result gives rise to an essential question: at
which moment does the decoherence occur?

In the above experiment, the strength of quantum entanglement between
the object state (electron in the AB ring) and the detector state
(electrons passing the QPC) is unknown due to the spatial separation.
Hence it is desirable to examine a system with strong entanglement
between an object and a detector.  In this Letter, we report
experiments on quantum dephasing, $i.e.$, phase randomization due to a
strong entanglement among spins in a QD and those of conducting
electrons.

\begin{figure}[]
\begin{center}
\includegraphics[scale=0.9]{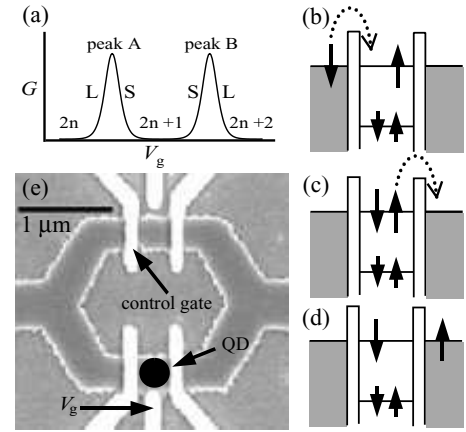}
\caption{(a) Illustration of two Coulomb oscillation peaks, labeled A
and B, and the expected magnitude of AB interference signal indicated
by ``L'' (large) and ``S'' (small). Because the parity of the number
of electrons in the QD changes by turns for successive Coulomb peaks
with changing gate voltage $V_{\rm g}$, the direction of the asymmetry
is expected to alternate. (b)-(d) Schematic drawings of energy diagram
of a QD and the dephasing process of a traversing electron by flipping
its spin in the QD (see text). (e) Scanning electron micrograph that
shows the sample geometry.}
\label{Fig1}
\end{center}
\end{figure}

For a system consisting of an AB ring and a QD, the possibility of
quantum decoherence due to strong object-detector coupling has been
pointed out theoretically~\cite{Akera,Konig}. The scenario is as
follows. Consider a QD where the single-electron energy levels with
Kramers degeneracy are distributed with nearly equal spacing according
to the random matrix theory (RMT)~\cite{mehta}.  Due to the
single-electron charging energy $E_{\rm c}$, electronic states of the
QD can be labeled with the electron number $N$.  Here we trace the
process where the ground state shifts as $|N=2n ({\rm even})\rangle$
$\rightarrow$ $|2n+1({\rm odd})\rangle $ $\rightarrow$ $|2n+2 ({\rm
even})\rangle$ with the increase in the gate voltage ($V_{\rm g}$) of
the QD.  There are two Coulomb peaks A and B in this process
corresponding to the two transitions, as schematically depicted in
Fig.~\ref{Fig1}(a).

On the left side of peak B, an electron passes the dot with the cycle
$|2n+1\rangle$ $\rightarrow$ $|2n+2\rangle$ $\rightarrow$
$|2n+1\rangle$.  In the initial state, the topmost orbital is
half-occupied with an electron of, for example, up-spin, as shown in
Fig.~\ref{Fig1}(b).  From the intermediate state (Fig.~\ref{Fig1}(c)),
an electron with either up- or down-spin can escape to the electrode.
If we label the spin states in the dot by $\vert d\uparrow\rangle$ (or
$\vert d\downarrow\rangle$) and those of the conducting electron by
$\vert c\uparrow\rangle$ (or $\vert c\downarrow\rangle$), the state
with an outgoing electron is written as $(\vert
d\downarrow\rangle\vert c\uparrow\rangle +\vert d\uparrow\rangle\vert
c\downarrow\rangle)/\sqrt{2}$ (Fig.~\ref{Fig1}(d)).  In this entangled
state, the spin-flip part $\vert d\uparrow\rangle\vert
c\downarrow\rangle$ is directly related to dephasing because it leaves
a trace on the QD~\cite{Konig}.

On the right side of peak B, the process changes to $|2n+2\rangle$
$\rightarrow$ $|2n+1\rangle$ $\rightarrow$ $|2n+2\rangle$ where no
spin-flip is allowed due to the Pauli principle.  Hence the total
quantum coherence is expected to be retained more on the right side
than on the left side of peak B.  The tendency on the left and right
sides of a Coulomb peak is reversed when $V_{\rm g}$ passes through
peak A. As a result, the AB amplitude changes as labeled by ``S"
(small) and ``L" (large) in Fig.~\ref{Fig1}(a).  An experimental
observation of such asymmetry~\cite{GefenCM2002}, therefore, provides
proof of ``partial coherence"~\cite{Konig} of the dot-ring
spin-entangled state.

We fabricated a QD-AB-ring system from a GaAs/AlGaAs two-dimensional
electron gas wafer (mobility $90~{\rm m^2/Vs}$ and sheet carrier
density $3.8 \times 10^{15}~{\rm m^{-2}}$) by electron-beam
lithography, wet etching and vacuum deposition of metallic gates
(Fig.~\ref{Fig1}(e)). A QD was formed in the lower arm of the AB ring
by negatively biasing two outer gates. The middle gate ($V_{\rm g}$)
was used to control the electrostatic potential of the QD. One of the
three gates on the upper arm was used to control the transmission of
the reference arm. The sample was cooled in a dilution refrigerator
with a base temperature of 30~mK. The conductance was measured by the
standard lock-in technique in a two-terminal setup.

The scenario for detecting partial coherence associated with the
spin-flip process described above is highly idealized in that all
aspects of the system and environment, other than the occupation of
the topmost level, are assumed to be identical throughout the region
of Coulomb peaks A and B. In the actual experiment, it is crucial to
assess to what degree this condition is fulfilled. Indeed, there are
many factors that might affect the AB amplitude as a function of $N$,
such as a change in the electrostatic potential. Furthermore, the
simplest approximation of RMT rarely holds for semiconductor
QDs~\cite{Patel1998PRL}, and electron correlation can give rise to
high-spin states. Therefore, the simple picture that single-electron
orbital levels are sequentially occupied by spin-up and -down
electrons is far from reality.

Nevertheless, one can hope to find an energy window ($i.e.$, a gate
voltage) where the simplest ``spin-pair" model is a good
approximation: Only a single Kramers degenerate state should exist
just above a closed-shell many-electron state in the energy diagram.
Although such a spin-pair state rarely exists in semiconductor
QDs~\cite{Patel1998PRL,Luscher2001PRL,Lindemann2002PRB}, once it is
found, we can circumvent the above problems and attribute the
difference in the coherence to the spin entanglement, because the
conditions other than the spin state are ideally equal on both sides
of Coulomb peaks in this window.

A spin-pair state appears as twin neighboring Coulomb peaks (spin-pair
peaks). The conditions required for such twin peaks are as follows.
I) They should be identical in their magnetic field dependences of
their positions and heights. II) The above dependence should be
different from those of neighboring ones because the conductions at
neighboring peaks are through different single-electron orbital
states. III) The addition energy between the peaks is likely to be
smaller than those of neighboring ones because there should be no
contribution of orbital energy.  Note that condition II excludes
high-spin states.

\begin{figure}[]
\begin{center}
\includegraphics[scale=0.9]{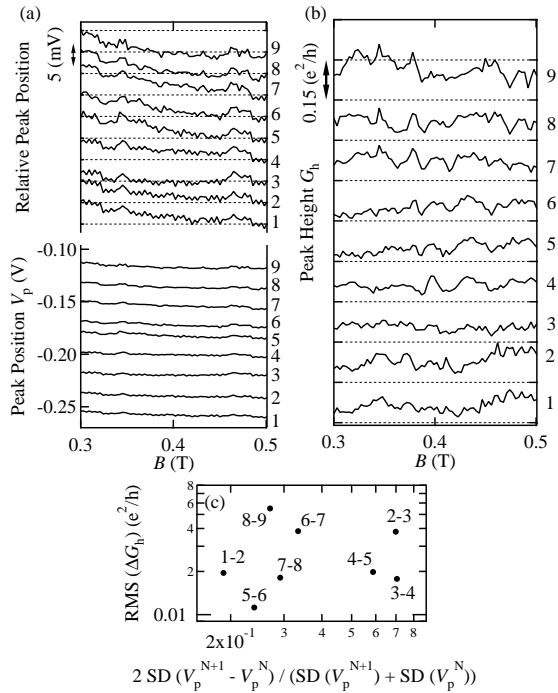}
\caption{Magnetic field dependence of (a) position and (b) height of
nine successive Coulomb peaks as a function of $B$. The peaks are
numbered as shown in the plot. The upper part of (a) shows the
relative shift of peak positions in the lower plot. (c) Root mean
square (RMS) value of the difference in peak height ($\Delta G_{\rm
h}$) in (b) plotted against the standard deviation (SD) of the
difference between $N+1$th and $N$th peak positions ($V_{\rm p}^{N+1}
- V_{\rm p}^{N}$) divided by the average value of each peak position
SD. The closer a point is to the lower left corner, the higher the
correlation is.}
\label{Fig2}
\end{center}
\end{figure}

With these criteria, we set out to find such twin Coulomb peaks.
Coulomb peaks are highly sensitive to environmental charge fluctuation
and a transition of a single-impurity around the sample changes their
positions.  On the other hand, the present experiment requires
complete stability throughout the measurement.  Periods for a single
set of measurements are thus limited and we can only find a single
pair which fulfills the above conditions, as described below.  The
positions ($V_{\rm p}$) and heights ($G_{\rm h}$) of nine successive
Coulomb peaks, which are obtained by fitting the standard formula of
the orthodox theory~\cite{Averin}, are shown in Figs.~\ref{Fig2} (a)
and (b), respectively, as a function of magnetic field $B$.  The
conductance of the reference arm is reduced to make the AB amplitude
small ($\sim$ 10~\% of $G_{\rm h}$), so the distribution of the
amplitude from peak to peak is negligible when we calculate the
correlation of the peak height.

As seen in Fig.~\ref{Fig2}(a), the line shapes of magnetic field
dependences of peak positions are similar.  This is natural in light
of the recent understanding of the nature of a wave function in
disordered quantum dots~\cite{Nakanishi}.  The essence of the theory
is as follows.  If we set the starting point at a dot with no
randomness, every orbital state has well-defined spatial symmetry. The
randomness of the confinement potential introduces ``children" states,
which are similar in spatial distribution to the parent state.  They
are different, however, particularly at the edge of the envelope, and
hence the difference should be emphasized by taking a close look at
the traces of position.  On the other hand, the conductance through
the dot is dominated by the edge of the wave function, and the Coulomb
peak height is more sensitive to the difference.  Actually, the
difference is clear in the peak heights shown in Fig.~\ref{Fig2}(b).

For a more quantitative comparison, we show in Fig.~\ref{Fig2}(c) the
RMS's of the difference in peak height ($\Delta G_{\rm h}$) and the
standard deviations (SDs) of the peak spacing ($V_{\rm p}^{N+1} -
V_{\rm p}^{N}$) divided by the average value of each peak position SD
$({\rm SD}(V_{\rm p}^{N+1}) + {\rm SD}(V_{\rm p}^{N}))/2$ for the
neighboring peak combinations in Fig.~\ref{Fig2}(a).  The pair 5-6
fulfills conditions I and III, {\it i.e.}, the point 5-6 is close to
the lower left corner and 4-5 and 6-7 are far from it.  This pair also
satisfies condition II, as shown in Fig.~\ref{Fig2}(a)~\cite{comment}.
From this observations we conclude that the pair 5-6 is a spin pair
which we have sought.

\begin{figure}[]
\begin{center}
\includegraphics{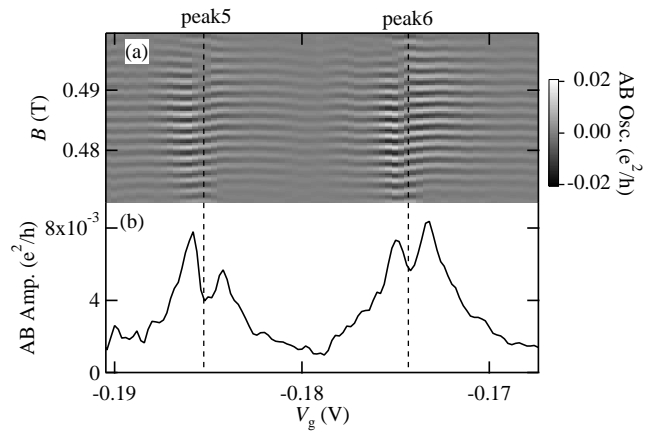}
\caption{(a) Gray-scale plot of the AB oscillation component against
the gate voltage $V_{\rm g}$ and the magnetic field $B$. Vertical
dashed lines represent the position of the Coulomb oscillation peaks
identified as a spin pair. (b) Averaged amplitude of AB oscillation
measured at each gate voltage. The AB amplitude is asymmetric with
respect to the center of the peak. The reduction of the amplitude at
the center of the peak is due to the $\pi$ jump of the AB phase
resulting from the two-terminal measurement geometry.}
\label{Fig3}
\end{center}
\end{figure}

We then proceed to the next step of examining the AB amplitude around
the spin-pair peaks 5 and 6.  Figure~\ref{Fig3}(a) is a gray-scale
plot of the AB component in the total conductance as a function of
$V_{\rm g}$ and $B$ extracted by fast Fourier transform.  We have
chosen the $B$ range on the basis of the stability of the peak
position in order to eliminate artifacts of level crossing, and the
peak positions are indicated by vertical dashed lines. As seen in
Fig.~\ref{Fig3}(a), the phase of the AB oscillation changes by $\pi$
at each Coulomb peak.  In some regions of the magnetic field, the
change is very steep and shows the feature of phase locking, reflecting
the two-terminal setup~\cite{Yeyati1995PRB,Yacoby1996PRB}.

In Fig.~\ref{Fig3}(b), we plot the AB amplitude averaged over ten
periods around $B = 0.485$~T as a function of $V_{\rm g}$. The large
dips in the AB amplitude at the peak position are due to phase locking
and separate the AB amplitudes to the left and right of the peaks,
which is an advantage of the two-terminal setup.  The two amplitude
peaks at both sides of each dip have different heights. For Coulomb
peak 5, the left peak is larger than the right, while the right peak
is larger than the left one for Coulomb peak 6.  This observation is
exactly what we expected from Fig.~\ref{Fig1}(a).  Note that possible
distortion due to the Fano effect is averaged out in the above
analysis because the Fano parameter oscillates sinusoidally with the
period of AB
oscillation~\cite{Kobayashi2002PRL,Kobayashi2003PRB,Konig}.  We
believe that all the possible artifacts are eliminated in the above
analysis, and our observation constitutes evidence for partial
coherence due to spin-flip scattering in a QD.  As for other
non-spin-pair peaks, we sometimes observed asymmetry, although their
direction seemed to change randomly.  We presume that the same spin
physics also plays a critical role in them, but cannot be conclusive
because the detailed information on the spin and orbital states at
those peaks is not known.

The shapes of the averaged amplitude in Fig.~\ref{Fig3}(b) for the two
Coulomb peaks are slightly different, while ideally, they would be
mirror images of each other with respect to the center of the Coulomb
valley. This is probably due to a remote effect of the gate electrode
to some part ({\it e.g.}, the reference arm) of the device other than
the quantum dot. Such a remote effect is linear in the gate voltage
(otherwise the total conductance should be largely affected by the
gate voltage) and only causes a distortion from peak to peak, thus,
the spin-pair approximation for the QD still holds.

For the spin-flip process, it is also presumed that the asymmetry
should be removed by applying a high magnetic field which lifts the
Kramers degeneracy~\cite{Akera,Konig}.  Unfortunately, in the present
experiment, such a high field was not attainable. However we have
performed the same procedure for about 50 Coulomb peaks of three
different samples at low ($\sim$ 0.5~T) and high ($\sim$ 2~T) fields;
the Zeeman energy of the latter is about 50~$\mu$eV for GaAs, and is
comparable to the spacing of the single-particle level in the QD
(typically $\sim 0.1~{\rm meV}$ under the present conditions).
Although no clear spin-pair state is found, at low fields, 80~\% of
the peaks exhibit the asymmetry of coherence, while 60~\% of them are
almost symmetric at high fields.  This suggests that the spin-flip
process plays an important role in the coherence through non-spin-pair
states, although, unfortunately, none of the measured peaks showed the
spin-pair features as clear as those of peaks 5-6.

We would like to comment on the question of whether spin flip or spin
entanglement is truly a dephasing process.  The diminishment of the AB
amplitude due to spin rotation is a coherent process and full rotation
to $4\pi$ recovers the original amplitude, as demonstrated by neutron
interference experiments~\cite{Werner1975PRL,Rauch1975PL}.  However,
in the system-plus-environment model, the spin rotation at a QD causes
diminishment of the coherence factor~\cite{Stern1990PRA,Zurek2003RMT},
if we classify the QD as part of the environment.  As mentioned in
Ref.~\cite{Konig}, when the Kondo state is fully developed, the
classification of the QD as part of the environment is invalid and the
dephasing due to the above spin scattering thus disappears at $T=0$.

In conclusion, we have observed the asymmetry of the AB interference
signal through a QD in an energy window for which the spin-pair state
is a good approximation.  The present results are in good agreement
with the theoretical predictions on quantum dephasing due to spin-flip
scattering and provide evidence of the partial coherence of electrons
which pass a QD with a localized moment.

We thank K.~Ensslin, Y.~Gefen, J.~K\"onig, and T.~Nakanishi for
helpful discussion.  This work is supported by a Grant-in-Aid for
Scientific Research and by a Grant-in-Aid for COE Research (``Quantum
Dot and Its Application'') from the Ministry of Education, Culture,
Sports, Science, and Technology of Japan. K.K. is supported by a
Grant-in-Aid for Young Scientists (B) (No.~14740186) from Japan
Society for the Promotion of Science.


\begin{thebibliography}{2}
\bibitem{comment} We cannot exclude the possibility that the pair 1-2
is also a spin-pair. Unfortunately the stability of the dot was not
enough to perform the measurement on this pair.
\end{thebibliography}
\end{document}